\documentclass[aps,11pt,onecolumn,preprintnumbers,amsmath,amssymb]{revtex4}

\usepackage[english]{babel}
\usepackage{graphicx}
\usepackage{color}

\begin{document}

\newcommand{\unit}[1]{\:\mathrm{#1}}            
\newcommand{\To}{\mathrm{T_0}}
\newcommand{\Tp}{\mathrm{T_+}}
\newcommand{\Tm}{\mathrm{T_-}}
\newcommand{\EST}{E_{\mathrm{ST}}}
\newcommand{\Rp}{\mathrm{R_{+}}}
\newcommand{\Rm}{\mathrm{R_{-}}}
\newcommand{\Rpp}{\mathrm{R_{++}}}
\newcommand{\Rmm}{\mathrm{R_{--}}}
\newcommand{\ddensity}[2]{\rho_{#1\,#2,#1\,#2}} 
\newcommand{\ket}[1]{\left| #1 \right>} 
\newcommand{\bra}[1]{\left< #1 \right|} 

\title{Valley Zeeman Effect in Elementary Optical Excitations of a Monolayer WSe2}
\author{Ajit Srivastava$^1$}
\author{Meinrad Sidler$^1$}
\author{Adrien V. Allain$^2$}
\author{Dominik S. Lembke$^2$}
\author{Andras Kis$^2$}
\author{A. Imamo\u{g}lu$^1$}
\affiliation{$^1$Institute of Quantum Electronics, ETH Zurich, CH-8093
Zurich, Switzerland}
\affiliation{$^2$Electrical Engineering Institute, Ecole Polytechnique Federale de Lausanne (EPFL), CH-1015
Zurich, Switzerland.}
\maketitle

*Correspondence to: imamoglu@phys.ethz.ch, sriva@phys.ethz.ch

\textbf{Abstract: A monolayer of a transition metal dichalcogenide (TMD) such
as WSe$_2$ is a  two-dimensional (2D) direct band-gap
valley-semiconductor~\cite{Mak10,Splendiani10} having an effective
Honeycomb lattice structure with broken inversion symmetry. The
inequivalent valleys in the Brillouin zone could be selectively
addressed using circularly-polarized light
fields~\cite{Cao12,Zeng12,Mak12}, suggesting the possibility for
magneto-optical measurement and manipulation of the valley
pseudospin degree of freedom~\cite{Xiao10,Xiao12,Xu14}. Here we
report such experiments that demonstrate the valley Zeeman effect --
strongly anisotropic lifting of the degeneracy of the valley
pseudospin degree of freedom using an external magnetic field. While
the valley-splitting measured using the exciton transition is
consistent with the difference of the conduction and valence band
orbital magnetic moments, the trion transition exhibits an
unexpectedly large valley Zeeman effect which cannot be understood
using an independent electron-hole picture. Instead, we find an
explanation using the recently predicted large Berry curvature and
the associated magnetic moment for the electron-hole exchange
interaction modified trion dispersion~\cite{Yu14}. Our results raise
the possibility of observing optical excitation induced valley Hall
effect in monolayer TMDs or topological states of photons strongly
coupled to trion excitations in a microcavity~\cite{Karzig14}.}\\
\textbf{Main Text} \\
Charge carriers in  two-dimensional (2D) layered materials with a
honeycomb lattice, such as graphene and TMDs, have a two-fold valley
degree of freedom labelled by $\pm K$-points of the Brillouin zone,
which are related to each other by time-reversal
symmetry~\cite{Xiao12}. In TMDs, the low energy physics takes place
in the vicinity of $\pm K$ points of the conduction and valence
bands with Bloch states that are formed primarily from $d_{z^2}$ and
$d_{x^2-y^2},d_{xy}$ orbitals of the transition metal,
respectively~\cite{Liu13}. The magnetic moment of charged particles
in a monolayer TMD arises from two distinct contributions: the
intra-cellular component stems from the hybridization of the
$d_{x^2-y^2}$ and $d_{xy}$ orbitals as $d_{x^2-y^2} \pm i d_{xy}$,
which provide the Bloch electrons at $\pm K$ in the valence band an
azimuthal angular momentum along $z$ of $l_z = \pm 2
\hbar$~(Fig.~1a). The second -- inter-cellular -- contribution
originates from the phase winding of the Bloch functions at $\pm
K$-points~\cite{Thonhauser05,Ceresoli06,Xiao07,Yao08}. In a two-band
model, this latter contribution to orbital magnetic moment is
identical for conduction and valence bands but much like the
intracellular contribution, is opposite in the two valleys.

In a 2D material such as a monolayer TMD, the current circulation
from the orbitals can only be within the plane; as a consequence, the
corresponding orbital magnetic moment can only point out-of-plane. A
magnetic field ($B$) along $z$ distinguishes the sense of
circulation in 2D, causing the opposite energy shifts ($-
\boldsymbol{\mu} \cdot \mathbf{B}$) in $\pm K$ valleys due to
opposite magnetic moments. The lifting of degeneracy between the two
valleys in presence of $B$, represents a valley analogue of the spin
Zeeman effect. Optical absorption or emission experiments
would allow for a direct determination of this valley Zeeman effect
since the valley index ($\pm K$) for independent electron-hole
pairs is linked to the helicity of light ($\sigma^{\pm}$) emitted
normal to the monolayer~\cite{Cao12,Zeng12,Mak12}. It is well known,
on the other hand, that the optical excitation spectra of TMDs are
strongly modified by Coulomb interactions, leading to strongly bound
neutral and charged exciton resonances. Moreover, it has been
recently shown that the electron-hole exchange interaction couples
the $\pm K$ valleys and results in exciton and trion energy
dispersion that is vastly different from the single-particle
description~\cite{Yu14}. It is therefore not a priori clear to what
extent the predictions about circular dichroism or orbital magnetic
moment that are based on a non-interacting particle picture remain
valid in optical measurements.

The samples studied in our experiments were obtained by mechanical
exfoliation of WSe$_2$ synthetic crystals onto heavily doped silicon
substrates with 285 nm SiO$_2$ layer on
top~\cite{Novoselov05,Radisavljevic11}. Monolayer flakes were
identified using their optical contrast. Polarisation-resolved photoluminescence (PL) and
resonant white-light reflection spectroscopy were performed in a
home-built confocal microscope setup placed in a liquid helium bath
cryostat. The sample temperature was 4.2 K and the excitation source
was a helium-neon (HeNe) laser at 632.8 nm or a tunable
continuous-wave (cw) Ti:Sapphire laser. The spot size for collection
wavelength was $\sim 1 \mu$m whereas that for 632.8 nm was $\sim$ 2-5
${\mu}$m. Magnetic fields in the range $\pm8.4$T were applied both
parallel and perpendicular to the plane of the sample. The polarisation control of excitation and photoluminescence
was done using a liquid crystal retarder calibrated for half and quarter wavelength retardance at exciton and trion wavelengths.

We perform polarisation-resolved photoluminescence (PL) spectroscopy on monolayer WSe$_2$ to identify the low energy optical excitations (see Methods). Figure~1b shows a typical polarisation-resolved PL spectra at zero
field. A sizeable ``valley coherence'' or linear dichroism ($\sim$
20\%) of the exciton peak (X$^{0}$) at $\sim$ 708 nm confirms its
monolayer nature~\cite{Jones13}. The peak at around 722 nm is
identified as originating from charged exciton (trion) (X$^{-}$)
emission, consistent with the previous PL studies on
WSe$_2$~\cite{Jones13}. Polarisation-resolved PL measurements
yielded a valley-contrasting circular dichroism of less than 20\%
for both X$^{0}$ and X$^{-}$ resonances in all the flakes that are
measured. As we argue below, the electron-hole exchange-induced
mixing of the two valleys is possibly responsible for reducing the
degree of circular dichroism~\cite{Yu14}.

Figure~2a shows the behaviour of  X$^{0}$ at various $B$ up to 8.4 T
in the Faraday geometry ($B \parallel z$). The PL is analysed in a
circularly polarised basis while the excitation of the laser is kept
linearly polarised, detuned by over 200 meV from the $X^0$
resonance. A clear splitting of the X$^{0}$ peak as a function of B
is observed: the splitting increases linearly with $B$ with slope of
about 0.25 meV/T (Fig.~2b). The magnitude of $B$-dependent splitting
in 5 different flakes are observed to be within $\pm 10 \%$ of this
value. We have also carried out resonant reflection measurements
showing a similar yet slightly smaller splitting of $\sim$ 0.21
meV/T for the exciton line (Fig.~2c). The fact that we do not
observe the trion peak in reflection measurements indicate that our
sample has a low doping density.

The measurement of the magnetic-field dependence of PL in Voigt
geometry ($B \perp z$) shows no observable splitting up to the
highest $B$ (Fig.~3a). This extreme anisotropy in the magnetic
response of the monolayer is a direct consequence of the fact that
the orbital magnetic moment of a strictly 2D material points
out-of-plane and thus can only couple to $B_z$. This observation
also rules out a spin-Zeeman contribution to the measured splitting.
Although the shift due to spin-Zeeman effect has opposite signs in
the two valleys, it is particle-hole symmetric in each valley and
therefore does not contribute to the splitting of optical
transitions, much like the intercellular orbital magnetic moment
(Fig.~2d). As expected, if the direction of $B$ in Faraday geometry
is reversed, the helicity of the split peaks also switches
(Fig.~3b).

When the B-dependent PL data is analysed in linear basis no clearly
resolvable splitting is observed (Fig.~3c); this is a consequence of
the fact that the split X$^{0}$ lines are circularly polarised and
have a linewidth that is larger than the maximum attainable B-field
induced splitting. By varying the polarisation setting of both the
excitation laser and the detection, it is confirmed that for finite
$B$ the split peaks are of opposite circular polarisation
irrespective of the polarisation of the excitation laser. Upon
linearly polarised excitation and circularly polarised detection,
the intensities of the two split peak is found to be about the same.

The observation of a splitting in the circularly-polarised basis
that scales linearly with the applied $B$ field is consistent with
an independent electron-hole description, as well as that of
electron-hole-exchange mediated coupling between the excitonic
excitations in the $\pm K$ valleys. In both models, the lifting of
the degeneracy occurs due to the equal but opposite orbital magnetic
moment of the two valleys together with unequal orbital magnetic
moments of conduction and valence bands within the same valley, as
shown in Fig~2d. For excitations with a vanishing in-plane momentum,
optical transition in each valley experiences a shift in energy
which is linear in $B$, given by $\delta E_{\pm K} =
-(\boldsymbol{\mu_c - \mu_v}) \cdot \mathbf{B} = (\mu_{\pm K,v} -
\mu_{\pm K,c}) B_z$ where $\mu_{-K,v(c)}$ = $- \mu_{K,v(c)}$ are the
sum of the intra- and inter-cellular orbital magnetic moments of the
valence (conduction) bands in the two valleys. Clearly, a necessary
requirement for observing splitting in optical transitions is
electron-hole symmetry breaking.

The magnitude of intra-cellular orbital magnetic moment  of valence
band in the two valleys is $\mu_v = (e/2m)l_z = 2 \mu_B$, where
$\mu_B = e\hbar/2m_e$ is the Bohr magneton. With this, one can
estimate the magnitude of the splitting arising purely from the
magnetic moment of the $d$-orbitals or the intra-cellular
contribution as $\Delta_{K,-K}^{intra} = \left(2 \mu_{B} -
(-2\mu_{B}) \right) B_{z} = 4 \mu_{B} B_{z}$. In a particle-hole
symmetric, two-band model, the values of inter-cellular orbital
magnetic moment are the same for conduction and valence band,
leading to a vanishing contribution to the splitting of optical
transition between the bands. TMDs break particle-hole symmetry and
a three-band model is needed to capture the band structure near the
$\pm K$ points as was recently shown~\cite{Liu13}. We calculate the
intercellular orbital magnetic moment for the two bands using the
three-band tight binding parameters for WSe$_2$~\cite{Liu13} and
find them to be comparable in magnitude to the intra-cellular
magnetic moment of the valence band (Supplementary Materials S1).
However, their difference is much smaller and should only have a
minor contribution to the splitting of optical transition. Indeed,
the observed splitting of about 4.3 $\mu_B B_z$ agrees well with the
intracellular contribution from the $d$-orbitals. Nevertheless, we
emphasize that in contrast to our experimental findings, the
three-band model suggests that the finite inter-cellular
contribution should reduce the value of the splitting to about  3.3
$\mu_B B_z$.


Figure~4a shows the polarisation dependent PL spectra of $X^{-}$ as
a function of $B$ in the Faraday geometry. Much like the $X^0$ peak,
a splitting of the two circularly polarised components of PL is
observed, which increases linearly with $B$ in Faraday geometry. As
in the exciton case, the Voigt geometry measurements did not yield
any measurable splitting. Surprisingly however, the magnitude of
$X^{-}$ splitting was measured to be significantly larger than that
of $X^{0}$ peak in all the flakes that were studied. As shown in
Fig.~4b, the splitting scales linearly with $B$ with a slope of
$\sim$ 0.32 - 0.36 meV/T which is more than 25$\%$  larger than that
of the $X^0$ peak.

A negatively (positively) charged trion is a three-body correlated
state comprising of a photo-generated electron-hole pair which binds
with an electron (hole) in the conduction (valence) band. The
binding energy of this three-body state is $\sim$ 30 meV, as
determined from the red-shift of $X^{-}$ peak from the $X^{0}$
resonance. The PL peak of $X^{-}$ arises from an optical transition
in which the initial state is that of the trion while the final
state has a Bloch electron in the conduction band.  Since
$B$-dependent PL splitting is determined by the difference in the
orbital magnetic moment between the initial and final states of the
optical transition, one expects the $B$-dependent $X^{0}$ and
$X^{-}$ PL splittings to be identical, since the extra electron in
the trion should contribute identically to the initial and the final
state magnetic moment. This simple non-interacting picture is
clearly inconsistent with our experimental findings and strongly
suggests the role of interactions in determining the magnetic moment
of trions.

Indeed, the degeneracy of the two trion states with center-of-mass
momentum $K$ ($-K$) that are comprised of an electron in $K$ ($-K$)
valley and a bright exciton occupying either the same or the
opposite valley, is lifted by the electron-hole exchange
interaction. The absence of valley coherence of $X^{-}$ peak in
monolayer WSe$_2$ has been attributed to the fast phase-rotation
associated with this energy splitting~\cite{Jones13}. Recently, it
was theoretically shown that the corresponding exchange-induced
modification of the trion dispersion ensures that the trion states
with center-of-mass momentum $\sim \pm K$ carry a large Berry
curvature~\cite{Yu14}. If the trion can be considered as a strongly
bound, charged quasi-particle, the exchange-induced Berry curvature
$\Omega (k)$ will be accompanied by a contribution to the orbital
magnetic moment that is given by $\mu (k) = \frac{e}{2\hbar} \Omega
(k) \delta_{ex}$ for $k \simeq \pm K$; here $\delta_{ex}$ is the
exchange-induced splitting of the trion resonances. Since this
additional contribution to the orbital magnetic moment exists only
in the initial state of optical emission, it could explain the
larger splitting of $X^{-}$ transition as compared to $X^{0}$ in our
experiments (Fig.~4c), provided that the $X^-$ PL originates mainly
from the lower trion branch where the exciton and the additional
electron occupy the same valley. In order to estimate the magnitude
of splitting based on this additional contribution to magnetic
moment, one needs to know the range of momenta of trion involved in
the radiative recombination and the Thomas-Fermi vector
corresponding to the carrier screening. In the absence of this
information, we refrain from making a quantitative estimate of the
expected trion splitting (Supplementary Materials S2). It should be
noted that this unique exchange-induced contribution to the magnetic
moment is very similar to the intercellular contribution in that it
stems from the non-trivial geometry of the trion band-structure;
however, it is optically induced and only exists within the lifetime
of trion.


Another expected experimental signature of the electron-hole
exchange interaction that acts as an exciton momentum-dependent
effective in-plane magnetic field for the valley
pseudospin~\cite{Yu14}, is a reduction of the circular dichroism.
An out-of-plane $B$ lifts the valley degeneracy and should, in
principle, overcome the valley-mixing due to exchange thereby
increasing the circular dichroism. Our measurements (Fig.~3d) show a
sizeable increase of the degree of circular dichroism by a factor of
1.7 - 2 at $B$ = 8.4 T as compared to its zero field value and are
qualitatively consistent with this prediction. However, a
quantitative estimation of this increase is not easy as it depends
on the details of relaxation processes following a non-resonant
excitation in a PL experiment.

While the  observation of valley Zeeman effect from the exciton
emission validates the predicted 2D band
structure of TMDs and establishes the valley degree of freedom as a pseudospin index; the anomalous $B$-dependent splitting of the
trion emission opens up exciting possibilities since it shows that
the elementary properties of these 2D semiconductors, such as the
magnetic moment and the Berry curvature could be modified by photon
absorption or emission. In addition, the exchange-induced Berry
curvature of trions should make the valley Hall
effect easier to measure as compared to an experiment based on excitons since no ionisation of optically generated electron-hole pair is needed~\cite{Mak14}. An interesting direction to pursue would be to couple a
TMD monolayer to the photonic modes of a high quality-factor cavity
and to use the exchange coupling of the trion states to create
topological states of trion-polaritons, as was recently proposed~\cite{Karzig14}.

During the preparation of the manuscript, we became aware of similar results by the University of Washington group~\cite{Aivazian14} and the Cornell group~\cite{MacNeill14}.
\\



{}

\vspace{1 cm}

\textbf{Acknowledgments} We acknowledge many enlightening discussions with Wang Yao, particularly regarding the role of intra-cellular current and the electron-hole exchange in optical excitations in TMDs. This work is supported by NCCR Quantum Science and
Technology (NCCR QSIT), research instrument of the Swiss National Science Foundation (SNSF)\\


%

\newpage

{\bf Figure 1: Photoluminescence of a monolayer WSe$_2$.}  {\bf a,} A schematic description of the
band structure showing the valley dependent intra-cellular
circulation of the valence band electrons. The double circulating clockwise (counter-clockwise) arrow denotes the out-of-plane orbital
magnetic moment of +2 $\mu_B$ (-2 $\mu_B$) arising from $d + i d$ ($d - i d$) orbitals in the valence band at $K$ ($-K$) valley.
The conduction band formed out of $d_{z^2}$ orbital carries no such magnetic moment. {\bf b,} A typical photoluminescence (PL) spectra at 4.2 K showing the
neutral exciton ($X^0$) and negatively charged exciton or trion
($X^{-}$) resonance at around 708 nm and 722 nm, respectively. The
laser polarisation is linear and the PL is detected co-polarised
(orange) or cross-polarized (purple) to it, showing linear dichroism
only for the $X^0$ peak.
\\

{\bf Figure 2: Magnetic field dependence of exciton
photoluminescence and the valley Zeeman effect} {\bf a,} Normalised
polarisation-resolved photoluminescence (PL) spectra of the neutral
exciton ($X^0$) peak as a function of the out-of-plane magnetic
field ($B$). The excitation laser is polarised linearly while the
red and blue traces correspond to PL analysed in circularly
polarised basis. The spectra at different $B$
are normalised to the zero field data while that at the same $B$ are
unnormalised. The solid lines are fit to the data. The $B$ dependent
splitting results from valley Zeeman effect as depicted in {\bf d}.
{\bf b,} Splitting extracted from fits in {\bf a} showing linear increase with $B$. Dashed line is a
linear fit to the data. {\bf c,} Resonant differential reflectance
spectra at +8.4 T  obtained using a filtered white light source.
Solid black lines are fits to the data using an admixture of
absorptive and dispersive lineshapes. {\bf d,} Valley
Zeeman effect - In a finite out-of-plane $B$, the degeneracy between
the $\pm$K valleys is lifted due to contributions from spin Zeeman
effect ($\Delta E_{\textrm{s}}$), the intercellular orbital magnetic
moment ($\Delta E_{\textrm{inter}}$), and the intracellular
contribution from the $d \pm i d$ orbitals of the valence band
($\Delta E_{\textrm{intra}}$). The signs of these contributions are
opposite in the two valleys. $\Delta E_{\textrm{s}}$ and $\Delta
E_{\textrm{inter}}$ for a 2-band model cause equal energy shifts of
the conduction and the valence band hence can not be detected in PL.
The resulting valley-split transitions are circularly
polarised.
\\

{\bf Figure 3: Strongly anisotropic magnetic response of
photoluminescence and its polarisation dependence.} {\bf a,} Exciton
photoluminescence spectra as a function of in-plane magnetic field
$B$ (Voigt geometry) showing no observable splitting even at the
highest applied field of 8.4 T. The dotted line depicts the emission
center wavelength which is identical for 0 T and 8.4 T. {\bf b,}
Polarity reversal of $B$ leads to switching of the emission helicity
of the two peaks together with a sign change of the splitting,
consistent with the valley Zeeman effect depicted in Fig.~2d. {\bf
c,} Polarisation dependence of magnetic splitting at 8.4 T showing no clear
splitting when the photoluminescence is analysed in the linear
basis, confirming that the split peaks are circularly polarised.
{\bf d,} Circular dichroism as a function of out-of-plane magnetic
field shows a clear increase by about a factor of 2. The
out-of-plane $B$ can overcome the valley-mixing due to electron-hole
exchange interaction leading to the observed increase.
\\

{\bf Figure 4: Magnetic field dependence of trion
photoluminescence.} {\bf a,} Normalised polarisation-resolved
photoluminescence (PL) spectra of the trion ($X^-$) peak as a
function of out-of-plane magnetic field ($B$). {\bf b,} Linear fit
to the splitting extracted from {\bf a} as function of $B$ gives a
larger value of the trion orbital magnetic moment as compared to the
exciton~(Fig.~2b). {\bf c,} An illustration of the different orbital
magnetic moments of initial $\ket{i}$ and final states $\ket{f}$ in $X^-$ optical
emission process. The blue (red) symbols label the $K$ ($-K$) valley. The electron-hole exchange energy ($\delta_{ex}$)
splits the trion dispersion into lower and upper branches with the excess electron and the exciton occupying the
same valley or the opposite. We assume that the radiative
recombination occurs predominantly from the same-valley branch with lower
energy. The initial state of the optical emission is thus a trion with exchange-induced magnetic moment while the final
state is a Bloch electron in the conduction band with an orbital
magnetic moment given by the intercellular circulation. The
difference between the two magnetic moments is finite and adds to
the intracellular contribution.
\\

\clearpage

\begin{figure}
\includegraphics[scale=0.8]{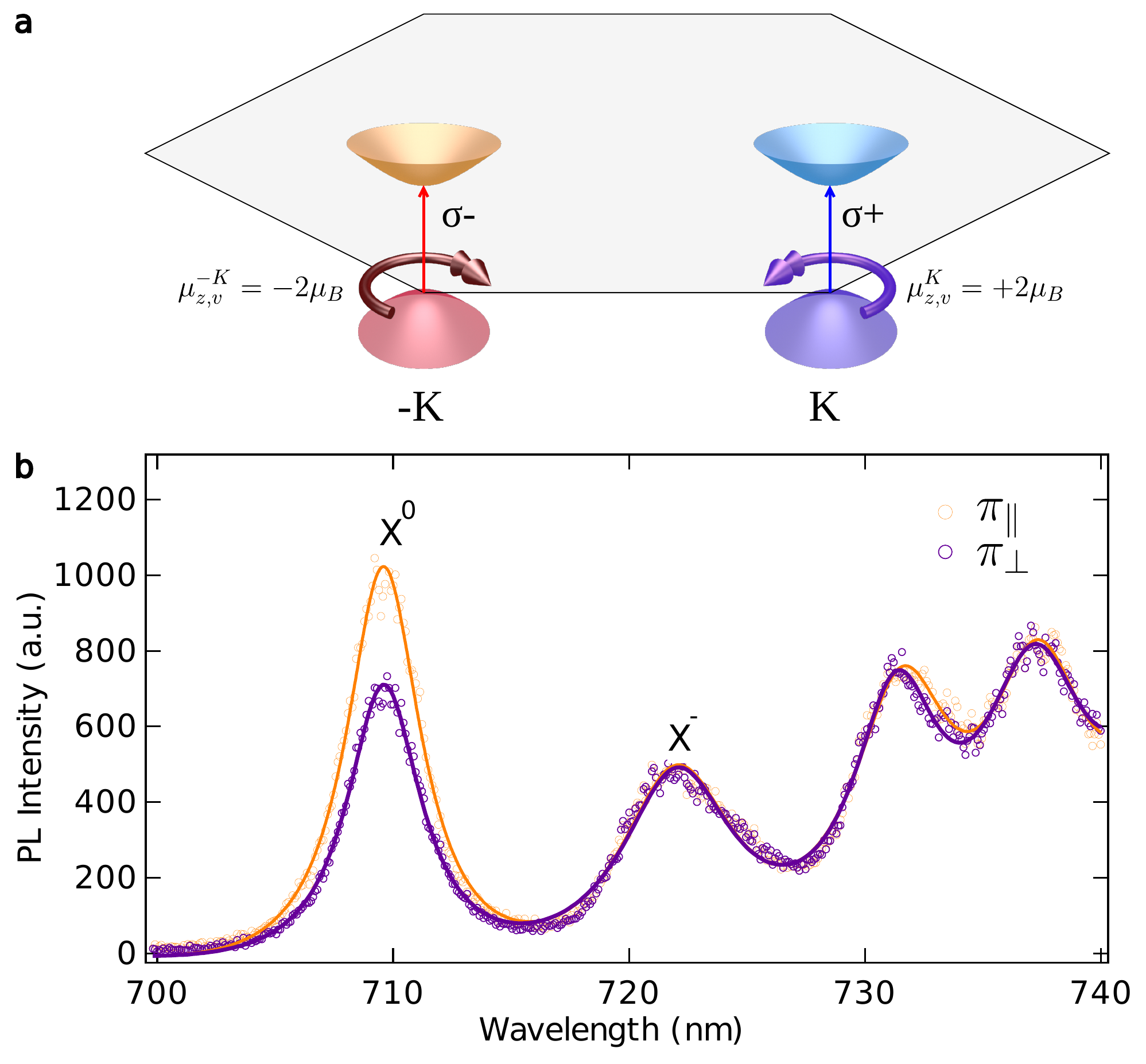}
\end{figure}

\clearpage

\begin{figure}
\includegraphics[scale=0.8]{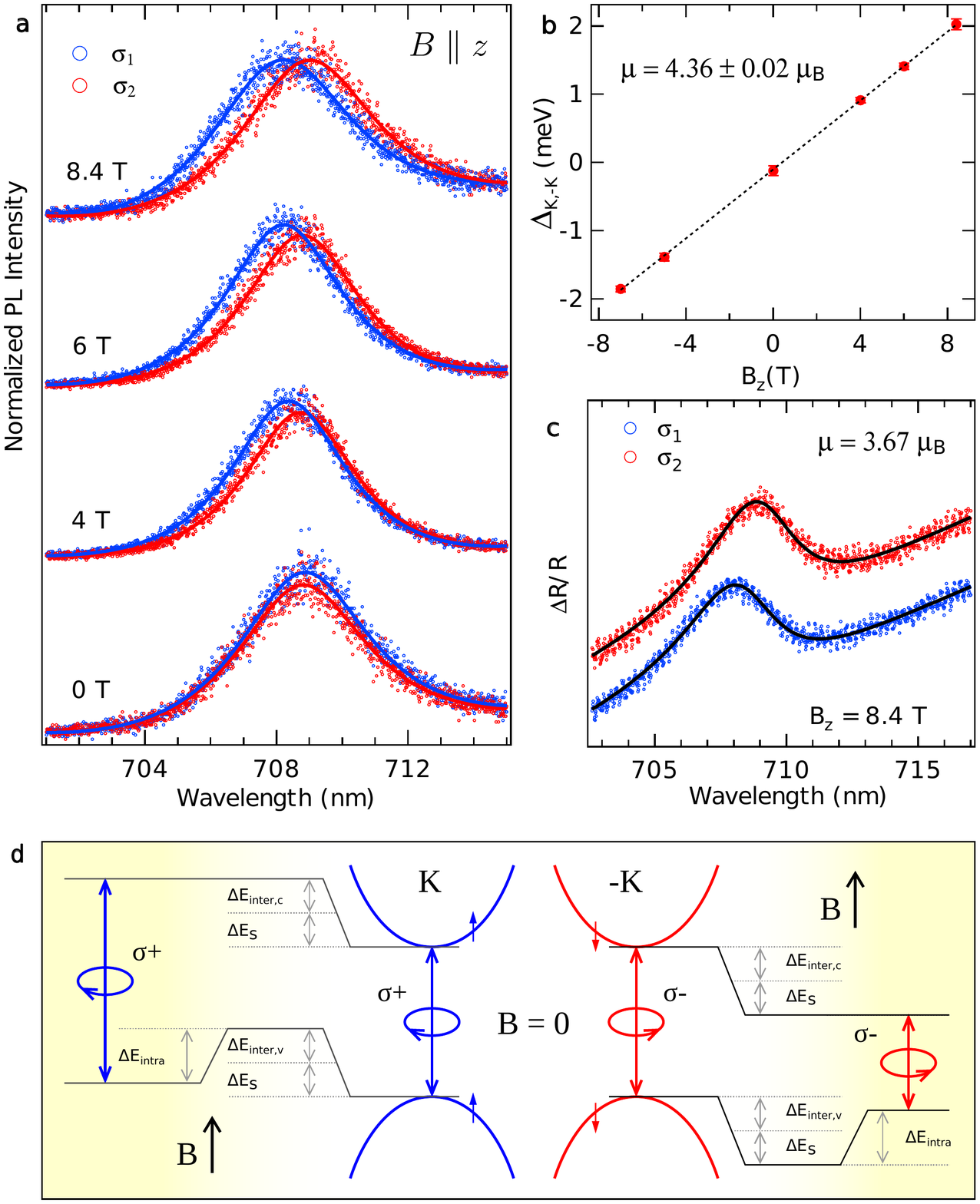}
\end{figure}

\clearpage

\begin{figure}
\includegraphics[scale=0.8]{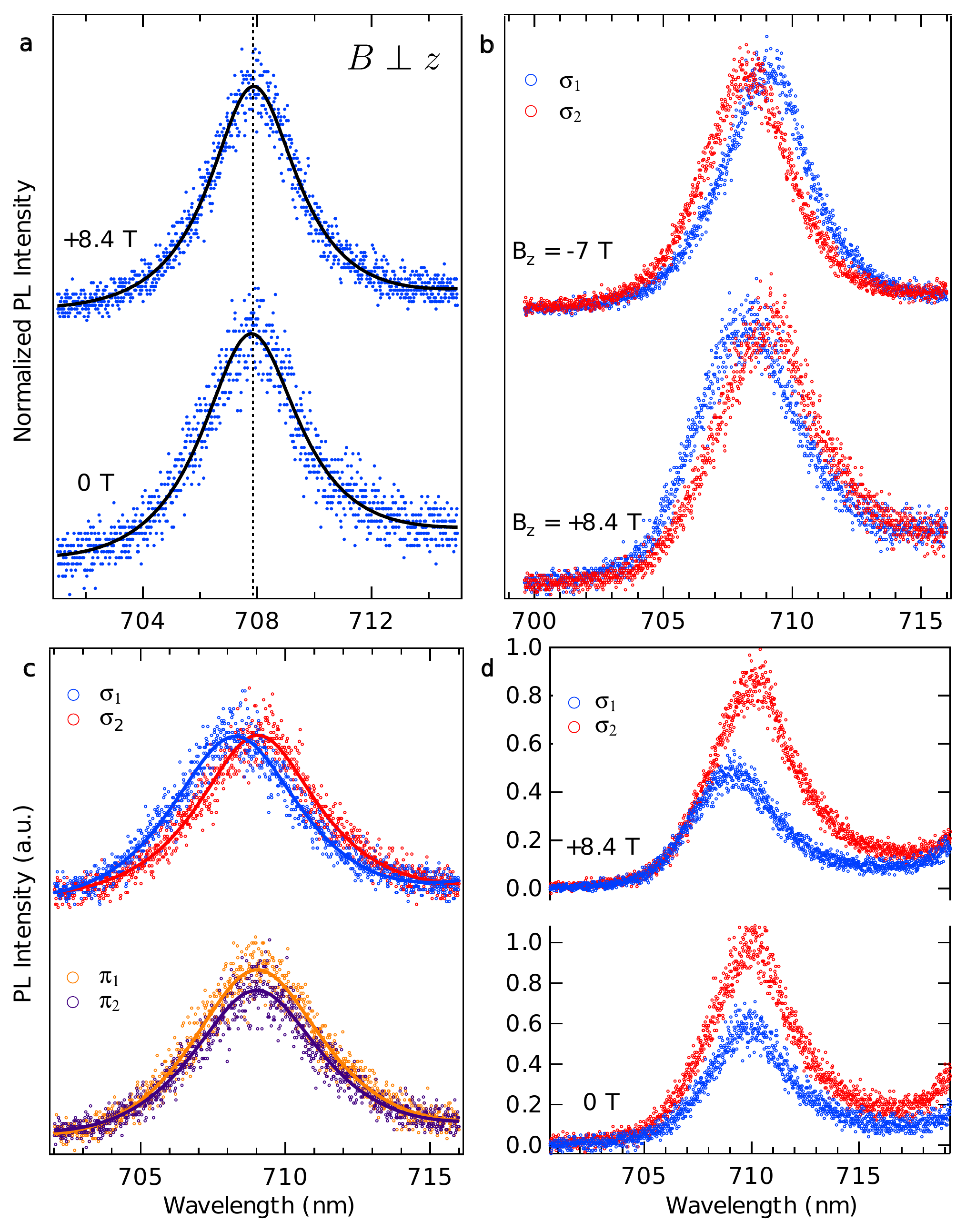}
\end{figure}

\clearpage

\begin{figure}
\includegraphics[scale=0.8]{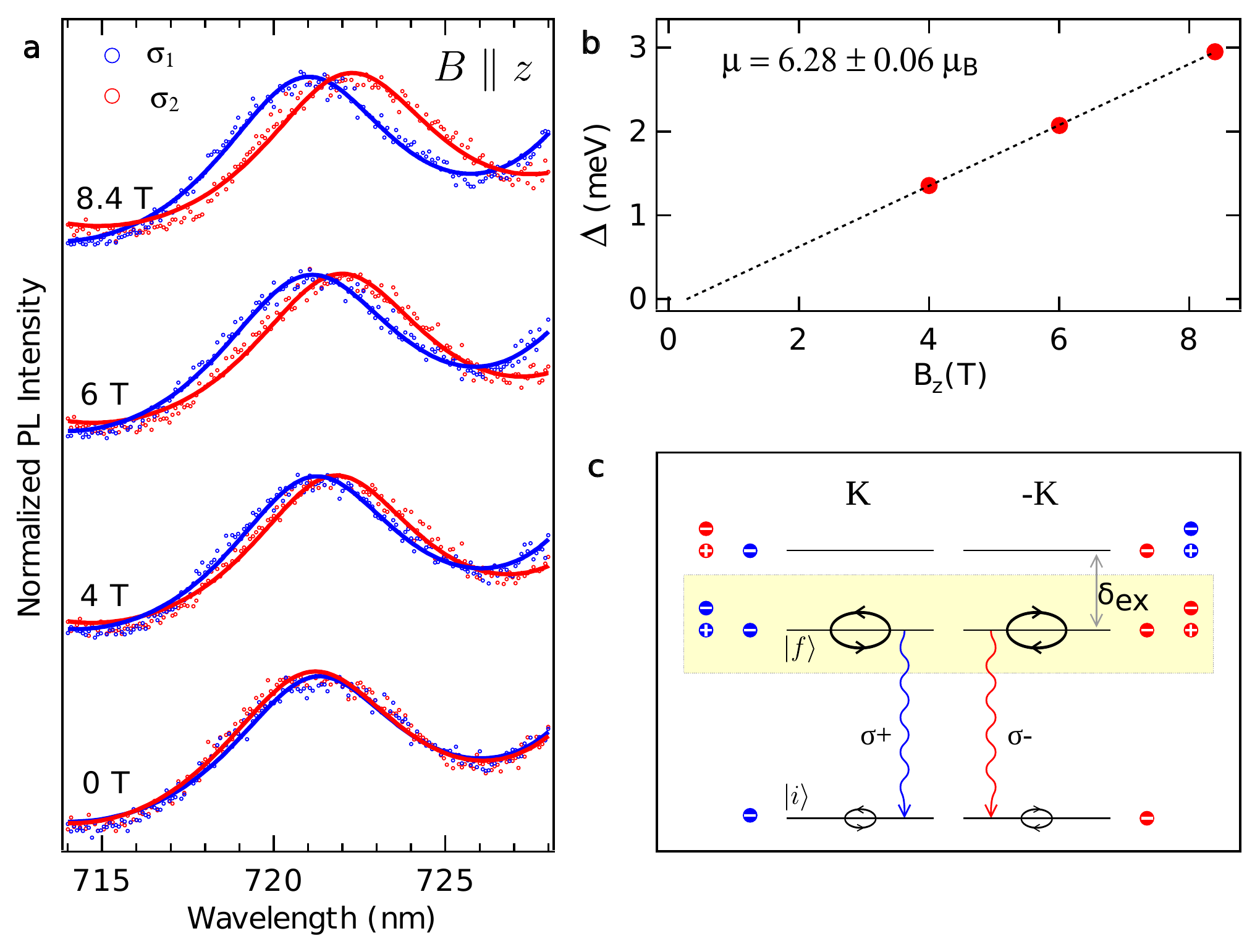}
\end{figure}

\clearpage

\textbf{Supplementary Information:}
\title{Valley Zeeman Effect in Elementary Optical Excitations of a Monolayer WSe2}
\author{Ajit Srivastava$^1$}
\author{Meinrad Sidler$^1$}
\author{Adrien V. Allain$^2$}
\author{Dominik S. Lembke$^2$}
\author{Andras Kis$^2$}
\author{A. Imamo\u{g}lu$^1$}
\affiliation{$^1$Institute of Quantum Electronics, ETH Zurich, CH-8093
Zurich, Switzerland}
\affiliation{$^2$Electrical Engineering Institute, Ecole Polytechnique Federale de Lausanne (EPFL), CH-1015
Zurich, Switzerland.}

\maketitle

\underline{\textbf{S1. Intercellular orbital magnetic moment in three-band tight binding model of WSe$_2$}}
As mentioned in the main text, the intercellular contribution to the orbital magnetic moment in a two-band model is identical for conduction and valence band. However, this particle-hole symmetry is broken in TMDs and one can expect a finite difference between the intercellular orbital magnetic moment of the conduction and the valence band. We use the three-band tight binding model of Liu \emph{et al.}~\cite{Liu13} to calculate the difference in orbital moments of the two bands. Eq.~4 of Liu \emph{et al.} publication presents a Hamiltonian which captures the band-structure of TMDs around $\pm K$-points $(\pm \frac{4\pi}{3a},0)$ - 
\begin{equation} \nonumber
{H}(\mathbf{k}) = \left( \begin{array}{ccc}
h_{0} & h_1 & h_2 \\
h_1^* & h_{11} & h_{12} \\
h_2^* & h_{12}^* & h_{22} 
\end{array} \right).
\end{equation}
We begin by expanding $H(k)$ at $K$-point to first order in $k_x$ and $k_y$, 
\begin{eqnarray}
h_0 &=& \epsilon_1 - 3t_0 \\
h_{1} &=& -\frac{3}{2}\left( \sqrt{3} t_2 k_y a + i t_1 k_x a \right) \\
h_{2} &=& \frac{3}{2}\left( \sqrt{3} t_2 k_x a + i t_1 k_y a \right) \\
h_{11} &=& \left( -\frac{3}{2} + \frac{3\sqrt{3}}{4}k_x a \right) t_{11} + \left( -\frac{3}{2} - \frac{3\sqrt{3}}{4} k_x a \right)t_{22} + \epsilon_2 \\
h_{22} &=& \left( -\frac{3}{2} + \frac{3\sqrt{3}}{4}k_x a \right) t_{22} + \left( -\frac{3}{2} - \frac{3\sqrt{3}}{4} k_x a \right)t_{11} + \epsilon_2 \\
h_{12} &=& \frac{3\sqrt{3}}{4} k_y a (t_{22} - t_{11}),
\end{eqnarray}
where the parameters $a$, $\epsilon_1$, $\epsilon_2$, $t_0$, $t_1$, $t_2$, $t_{11}$, $t_{22}$, and $t_{12}$ are to be taken from Table II of Liu \emph{et al}. It is convenient to make the above Hamiltonian traceless by choosing the energy datum as $\frac{1}{3}$Tr($H$). The Hamiltonian then reduces to -

\begin{equation}
\tilde{H}(k_x,k_y) = \left( \begin{array}{ccc}
-2c & -v a k_y - i u a k_x & v a k_x - i  u a k_y \\
-v a k_y + i u a k_x & c + d a k_x & -d a k_y - i w \\
v a k_x + i u a k_y & -d a k_y + i w & c - d a k_x 
\end{array} \right),
\end{equation}

where, 
\begin{eqnarray}
c &=& \frac{\left(\epsilon_2 - \epsilon_1\right)}{3} - \frac{1}{2}(t_{11} + t_{22} - 2t_0) \\
u &=& \frac{3}{2}t_1 \\
v &=& \frac{3\sqrt{3}}{2}t_2 \\
w &=& 3\sqrt{3} t_{12} \\
d &=& \frac{3\sqrt{3}}{4}(t_{11} - t_{22}).
\end{eqnarray}

At $K$-point, the eigenvalues are -
\begin{eqnarray}
E_1 &=& c-w \quad \textrm{valence band}\\
E_2 &=& -2c \quad \textrm{conduction band}\\ 
E_3 &=& c+w \quad \textrm{higher band}.
\end{eqnarray}  The corresponding eigenvectors are $u_1 = \frac{1}{\sqrt{2}}\left( 0, i , 1 \right)^T$, $u_2 = \left( 1, 0, 0 \right)^T$ and $u_3 = \frac{1}{\sqrt{2}}\left( 0, -i , 1 \right)^T$ which in the basis $\{ \ket{d_{z^2}}, \ket{d_{x^2 - y^2}}, \ket{d_{xy}} \} $ confirm that the valence band at $K$-point is formed out of $\frac{1}{\sqrt{2}} \left( \ket{d_{x^2 - y^2}} + i \ket{d_{xy}}\right)$ orbital while the conduction band is made solely from $\ket{d_{z^2}}$ orbital, within this three-band model. 

From Eq.~7, one easily calculates $\partial \tilde{H} / \partial k_x$ and $\partial{\tilde{H}} / \partial k_y$ at $K$ to be -
\begin{equation}
\frac{\partial \tilde{H}}{\partial k_x}\bigg |_\mathbf{K} = \left( \begin{array}{ccc}
0 &  - i u a & v a  \\
i u a &  d a & 0 \\
v a & 0 &  - d a 
\end{array} \right),
\end{equation} and
\begin{equation}
\frac{\partial \tilde{H}}{\partial k_y}\bigg |_\mathbf{K} = \left( \begin{array}{ccc}
0 & -v a & -i  u a \\
-v a & 0 & -d a \\
i u a & -d a & 0
\end{array} \right),
\end{equation}
The orbital angular momentum $L_n (k)$ for the n-th band is given by the following expression~\cite{Yafet63,Chang96},
\begin{equation}
L_n (k) = i \frac{m}{\hbar} \sum_{j \neq n} \left[ \frac{\bra{u_{n,k}} \frac{\partial H}{\partial k_x} \ket{u_{i,k}} \bra{u_{i,k}}  \frac{\partial H}{\partial k_y} \ket{u_{n,k}}}{E_i - E_n} - \textrm{c.c} \right],
\end{equation}
yielding -
\begin{eqnarray}
L_1(\mathbf{K}) &=& -\frac{m}{\hbar}a^2 \left( \frac{(u+v)^2}{3c-w} + \frac{d^2}{w} \right) \\
L_2(\mathbf{K}) &=& -\frac{m}{\hbar}a^2 \frac{2w(u^2+v^2)+6cuv}{(9c^2-w^2)}.
\end{eqnarray} The orbital magnetic moment for the n-th band is then $\mu_n(k) = \frac{e}{2m} L_n(k)$. Plugging the numbers for the parameters, we get -
\begin{eqnarray}
\mu_{v,inter} (\mathbf{K}) = 3.49~\mu_\textrm{B} \quad \textrm{(GGA)}\\
\mu_{c,inter} (\mathbf{K}) = 3.83~\mu_\textrm{B} \quad \textrm{(GGA)} \\
\mu_{v,inter}  (\mathbf{K}) = 3.41~\mu_\textrm{B} \quad \textrm{(LDA)}\\
\mu_{c,inter}  (\mathbf{K}) = 3.75~\mu_\textrm{B} \quad \textrm{(LDA)}
\end{eqnarray} The total orbital magnetic moment at $K$-point then becomes $\mu_{tot} = \mu_{inter} + \mu_{intra}$, yielding a splitting of 3.34 $\mu_\textrm{B}B_z$ (GGA) and 3.32 $\mu_\textrm{B}B_z$ (LDA).

\vspace{1cm}

\underline{\textbf{S2. Orbital magnetic moment of trion in WSe$_2$ due to finite Berry curvature}}

According to Yu \emph{et al.}~\cite{Yu14}, the trion ($X^-$) dispersion near $\pm K$-point has an exchange-induced gap of $\delta_{ex}$ which results in a large Berry curvature, $\Omega_{X^-} (k)$. If we treat $X^-$ as a charged quasi-particle described by a two-band model of Yu \emph{et al.} near $\pm K$-points, the finite Berry curvature is accompanied by an orbital angular momentum $L(k)$ which is identical for both bands. Using Eq.~18 of S1, we get $L(\mathbf{k})$ as,
\begin{equation}
L(\mathbf{k}) =  i \frac{m}{\hbar} \left[ \frac{\bra{u_{1,k}} \frac{\partial H}{\partial k_x} \ket{u_{2,k}} \bra{u_{2,k}}  \frac{\partial H}{\partial k_y} \ket{u_{1,k}}}{E_g (k)} - \textrm{c.c} \right].
\end{equation} 
where $E_g (k) = \delta_{ex}\left(1+\frac{4J^2k^4}{K^2 \delta_{ex}^2 (k+k_{TF})^2}\right)^{1/2} $ is the $k$-dependent gap. Expressing in terms of $\Omega_{X^-}(k)$, 
\begin{equation}
L(k) = \frac{m}{\hbar} E_{g}(k) \Omega_{X^-}(k). 
\end{equation} 
The trion magnetic moment is then given by $\mu_{X^-} = \frac{e}{2m} L(k)$, 
\begin{equation}
\mu_{X^-}(k) = \frac{eJ^2}{\hbar K^2 \delta_{ex}}\frac{k^2(k+2k_{TF})}{(k+k_{TF})^3}\left(1+ \frac{4k^4J^2}{(k+k_{TF})^2 K^2 \delta_{ex}^2} \right)^{-1}
\end{equation}
where $J$ is the electron-hole exchange coupling strength and $k_{TF}$ is the Thomas-Fermi wavevector corresponding to carrier screening. Fig.~S1(a) shows $\mu(k)$ in units of $\mu_B$ for $k$ up to 0.01$K$ and three different values of $k_{TF}$. A $k_{TF}$ of 0.1 $\omega_0/c$ corresponds to a charge density of $\sim$ 10$^9$ cm$^{-2}$.

The average (intercellular) magnetic moment depends on the range of $k$-wavevectors involved in the radiative recombination and the carrier doping density through $k_{TF}$. It can be calculated by averaging over certain wave vector range $\delta k$, say from 0 to $k_{lim}$. This is shown in Fig.~S1(b) for different values of $k_{TF}$ and different values of $k_{lim}$ (from $k_{ph} = \omega_0/c$ to $0.03K$).
 
In order to observe a splitting of 5.5 to 6.2 $\mu_B$ the trion final state intercellular magnetic moment should be between 4.55 and 4.9 $\mu_B$ since the intracellular contribution to orbital magnetic moment is $\mu_{intra}$ = 2$\mu_B$, and the intercellular magnetic moment for the initial state of an electron in the conduction band was estimated to be $\sim$ 3.8 $\mu_B$ (see supplementary S1). This interval of $\mu_{avg}$ is shown as white band in Fig.~S1(b). It is evident that a wide of range of parameters can lead to the experimentally observed trion splitting.
 
{}

\newpage

{\bf Figure S1: Orbital magnetic moment of trion arising from large Berry curvature.}  {\bf a,} Momentum-resolved orbital magnetic moment of trion ($X^-$) as a function of $X^-$ centre of mass wavevector $k$ measured from the $K$-point for different Thomas-Fermi wave vectors ($k_{TF}$) of carrier doping densities. {\bf b,} Average magnetic moment of $X^-$ for a range of wavevectors involved in radiative recombination ($\delta k$) and $k_{TF}$. The white band depicts the range of magnetic moments consistent with the experimentally observed trion splitting.
\\

\clearpage

\begin{figure}
\includegraphics[scale=0.8]{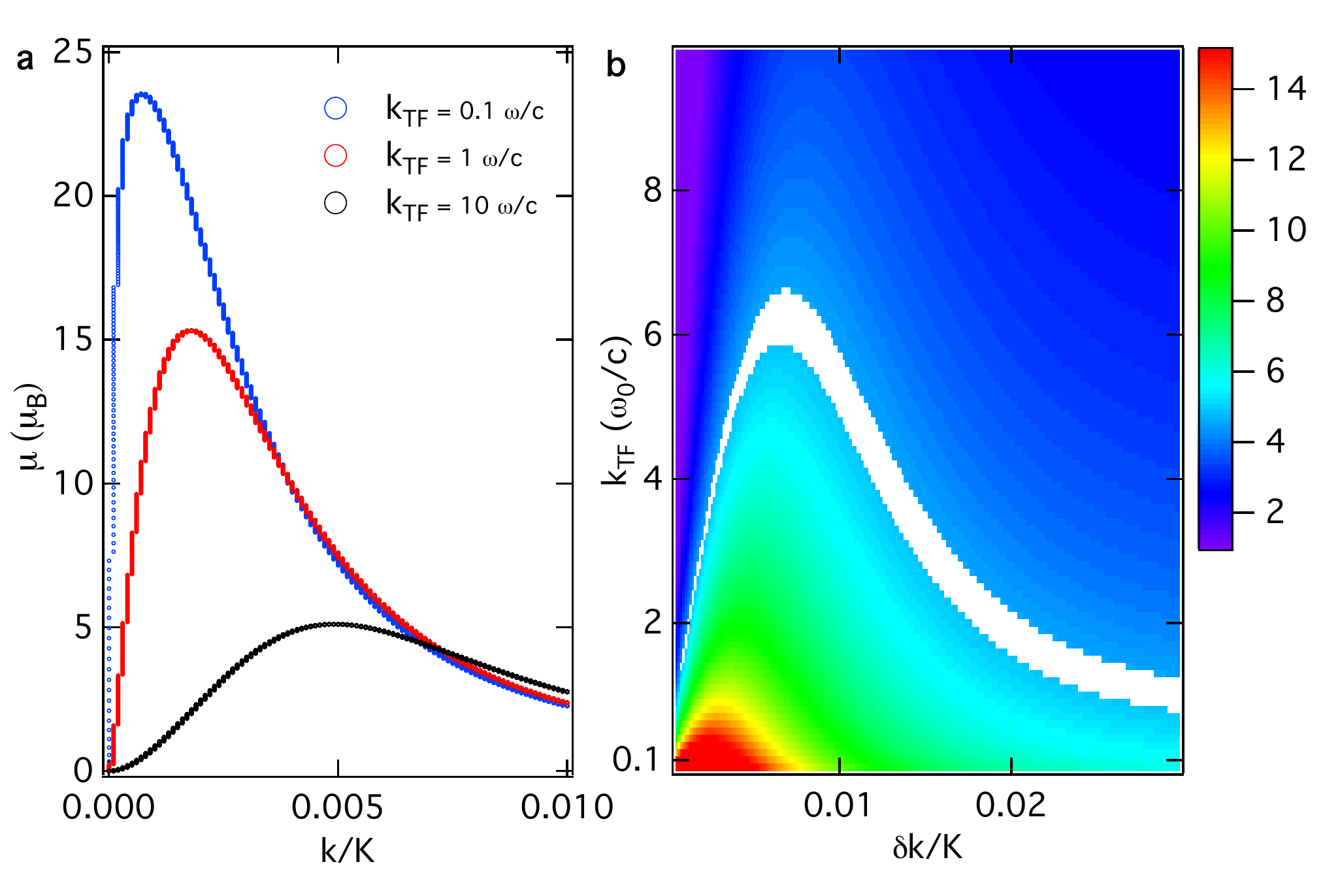}
\end{figure}

\end{document}